\documentstyle[12pt]{article}

\global\arraycolsep=2pt
\makeatletter
\def\fmslash{\@ifnextchar[{\fmsl@sh}{\fmsl@sh[0mu]}}
\def\fmsl@sh[#1]#2{%
  \mathchoice
    {\@fmsl@sh\displaystyle{#1}{#2}}%
    {\@fmsl@sh\textstyle{#1}{#2}}%
    {\@fmsl@sh\scriptstyle{#1}{#2}}%
    {\@fmsl@sh\scriptscriptstyle{#1}{#2}}}
\def\@fmsl@sh#1#2#3{\m@th\ooalign{$\hfil#1\mkern#2/\hfil$\crcr$#1#3$}}
\makeatother

\begin{document}

\vspace{1.cm}
\begin{center}
\Large\bf Couplings of Pions with Heavy Baryons From\\ 
Light-Cone QCD Sum Rules\\
 in the Leading Order of HQET 
\end{center}
\vspace{0.5cm}
\begin{center}
{Shi-lin Zhu and Yuan-Ben Dai}\\\vspace{3mm}
{\it Institute of Theoretical Physics,
 Academia Sinica, P.O.Box 2735, Beijing 100080, China }
\end{center}

\vspace{1.cm}
\begin{abstract}
 The couplings of pions with heavy baryons  
$g_2(\Sigma^* , \Sigma )$ and $g_3(\Sigma^* , \Lambda )$
are studied with light-cone QCD sum rules in the leading order 
of heavy quark effective theory. Both sum rules are stable. Our 
results are $g_2=1.56\pm 0.3\pm 0.3$, $g_3=0.94\pm 0.06\pm 0.2$.
\end{abstract}



\vspace{1.cm}

\pagenumbering{arabic}

\section{ Introduction}
\label{sec1} 
Important progress has been achieved in the interpretation of 
heavy hadrons composed of a heavy quark with the development 
of the heavy quark effective theory (HQET) \cite{grinstein}. HQET provides a 
systematic expansion of the heavy hadron spectra and transition amplitude
in terms of  $1/ m_Q$, where $m_Q$ is the heavy quark mass.
Of course one has to employ some specific 
nonperturbative methods to arrive at the detailed predictions. Among the
various nonpeturbative methods, QCD sum rules is useful to extract the 
low-lying hadron properties \cite{svz}. 

The couplings of the heavy mesons with pions has been analysed 
with QCD sum rules [5-15]. The couplings of heavy baryons with 
soft pions are estimated from QCD sum rules in an 
external axial field \cite{grozin97}.
In this approach the mass difference $\Delta$ between the baryons in the 
initial and final states is approximately taken to be zero.
 
In this work we employ the light-cone QCD sum rules (LCQSR)
in HQET to calculate the couplings $g_{2,3}$ 
to the leading order of $1/ m_Q$.
The LCQSR is quite different from the conventional QCD sum rules, which is 
based on the short-distance operator product expansion (OPE). The 
LCQSR is based on the OPE on the light cone, 
which is the expansion over the twists of the operators. The main contribution
comes from the lowest twist operator. Matrix elements of nonlocal operators 
sandwiched between a hadronic state and the vacuum defines the hadron wave
functions. The LCQSR approach has the advantage that 
the double Borel transformation is used so that the  
the continuum contribution is treated  in a way better than the external 
field approach. Moreover, the final 
sum rule depends only on the value of the wave function at a specific point like
$\varphi_{\pi}(u_0 ={1\over 2})$, which is much better known than the whole wave 
function \cite{bely95}.

%%%%%%%%%%%%%%%%%%%%%%%%%%%%%%%%%%%%%%%%%%%%%%%%%%%%%%%%%%%%%%%%%%%%%%%%%%%%%%%%

\section{ Sum rules for the coupling constants}

We first introduce the interpolating currents for the heavy baryons:
\begin{equation}
\label{lam1}
\eta_{\Lambda} (x) =\epsilon_{abc} [{u^a}^T(x) C\gamma_5 d^b (x)] h_v^c (x) \, ,
\end{equation}
\begin{equation}
\label{sig1}
\eta_{\Sigma^+} (x) =\epsilon_{abc} [{u^a}^T(x) C\gamma_{\mu} d^b (x)] 
\gamma^{\mu}_t \gamma_5 h_v^c (x) \, ,
\end{equation}
\begin{equation}
\label{star1}
\eta^{\mu}_{{\Sigma^{++}}^*} (x) =\epsilon_{abc} [{u^a}^T(x) C\gamma_{\nu} u^b (x)] 
(-g_t^{\mu\nu}+ {1\over 3} \gamma_t^{\mu}\gamma_t^{\nu} )h_v^c (x) \, ,
\end{equation}
where $a$, $b$, $c$ is the color index, $u(x)$, $d(x)$, $h_v(x)$ is the up,
down and heavy quark fields, $T$ denotes the transpose, $C$ is the charge conjugate 
matrix,
$g_t^{\mu\nu}=g^{\mu\nu}-v^{\mu}v^{\nu}$, $\gamma_t^{\mu}=\gamma_{\mu}-
{\hat v}v^{\mu}$, and $v^{\mu}$ is the velocity of the heavy hadron.

The overlap amplititudes of the interpolating currents with the heavy baryons 
is defined as:
\begin{equation}
\label{lam2}
\langle 0|\eta_{\Lambda} |\Lambda\rangle =f_{\Lambda}u_{\Lambda} \, ,
\end{equation}
\begin{equation}
\label{sig2}
\langle 0|\eta_{\Sigma} |\Sigma\rangle =f_{\Sigma} u_{\Sigma}\, ,
\end{equation}
\begin{equation}
\label{star2}
\langle 0|\eta^{\mu}_{\Sigma^*} |\Sigma^*\rangle ={f_{\Sigma^*}\over \sqrt{3}}
 u^{\mu}_{\Sigma^*} \, ,				
\end{equation}
where $u^{\mu}_{\Sigma^*}$ is the Rarita-Schwinger spinor in HQET. 
In the leading order of HQET, $f_{\Sigma}= f_{\Sigma^*}$ \cite{mass}.

We adopt the same notations for $g_{2,3}$ as in \cite{cho}. 
The coupling constants
$g_2$ and $g_3$ are defined through the following amplitudes:
\begin{equation}
\label{coup-g3}
M(\Sigma^*_c\to \Lambda_c \pi )=i{g_3\over f_\pi} {\bar u}_{\Lambda_c} q_\mu
u^\mu_{\Sigma^*_c} \; ,
\end{equation}
\begin{equation}
\label{coup-g2}
M(\Sigma^*_c\to \Sigma_c \pi )={g_2\over \sqrt{6}f_\pi} 
i\epsilon_{\nu \rho \sigma \mu} v^\sigma q_t^\nu {\bar u}_{\Sigma_c} 
\gamma_t^\rho \gamma_5 u^\mu_{\Sigma^*_c} \; ,
\end{equation}
where $f_\pi =132$MeV, $q_\mu$ is the pion momentum. The process (\ref{coup-g2}) 
is kinematically forbidden. It is very important to get a reliable estimate 
of the coupling $g_2$ since it is not directly accessible experimentally. 

In order to derive the sum rules for the coupling constants we consider 
the correlators 
\begin{equation}
\begin{array}{ll}
\label{sig4}
 \int d^4x\;e^{-ik\cdot x}\langle\pi(q)|T\left(\eta^{\mu}_{\Sigma^*}(0)
{\bar \eta}_{\Sigma}(x)\right)|0\rangle = &
 {1+{\hat v}\over 2}(-g_t^{\mu\nu}+{1\over 3}\gamma_t^{\mu}\gamma_t^{\nu}) 
 \\
  & i\epsilon_{m\rho n\nu} q_t^m v^n \gamma_t^{\rho}\gamma_5
  G_{\Sigma^* ,\Sigma}(\omega,\omega')\;,
\end{array}
\end{equation}
\begin{equation} 
\label{lam4}
 \int d^4x\;e^{-ik\cdot x}\langle\pi(q)|T\left(\eta^{\mu}_{\Sigma^*}(0)
 {\bar \eta}_{\Lambda}(x)\right)|0\rangle =	
   {1+{\hat v}\over 2}q_t^{\nu} (-g_t^{\mu\nu}+{1\over 3}\gamma_t^{\mu}\gamma_t^{\nu})\\
   G_{\Sigma^* ,\Lambda}(\omega,\omega')\;,
\end{equation}
where $k^{\prime}=k-q$, $q^t_\mu =q_\mu -(q\cdot v) v_\mu$, 
$\omega=2v\cdot k$, $\omega^{\prime}=2v\cdot k^{\prime}$ and $q^2=0$. 

Let us first consider the function $G_{\Sigma^* ,\Lambda}(\omega,\omega^{\prime})$ in (%
\ref{lam4}). As a function of two variables, it has the following pole terms
from double dispersion relation 
\begin{eqnarray}
\label{pole}
{4g_3 \over \sqrt{3} f_\pi}{f_{\Sigma^*}f_{\Lambda}\over (2\bar\Lambda_{\Sigma^*}
-\omega')(2\bar\Lambda_{\Lambda}-\omega)}+{c\over 2\bar\Lambda_{\Sigma^*}
-\omega'}+{c'\over 2\bar\Lambda_{\Lambda}-\omega}\;,
\end{eqnarray}
where $f_{\Sigma^*}$ etc are constants defined in 
(\ref{lam2})-(\ref{star2}), $\bar\Lambda_{\Sigma^*}=m_{\Sigma^*}-m_Q$. 

Neglecting the four particle component of the pion wave function, 	
the expression for $G_{\Sigma^* , \Lambda}(\omega, \omega')$ 
with the tensor structure reads
\begin{equation}\label{lam5}			
4  \int_0^{\infty} dt \int dx e^{-ikx} 
\delta (-x-vt){\bf Tr} \{ \langle \pi (q)|u(0) {\bar d}(x) |0\rangle  
 [\gamma_5 C iS^T(-x) C\gamma_{\nu}] \} \; ,
\end{equation}
where $iS(-x)$ is the full light quark propagator with both perturbative  
term and contribution from vacuum fields. 
\begin{eqnarray}\label{prop}\nonumber
iS(x)=\langle 0 | T [q(x), {\bar q}(0)] |0\rangle &\\
=i{{\hat x}\over 2\pi^2 x^4} -{\langle {\bar q} q\rangle  \over 12}
-{x^2 \over 192}\langle {\bar q}g_s \sigma\cdot G q\rangle &\\ \nonumber
-ig_s{1\over 16\pi^2}\int^1_0 du \{
{{\hat x}\over x^2} \sigma\cdot G(ux)-4iu {x_\mu\over x^2} 
G^{\mu\nu}(ux)\gamma_nu \} +\cdots  & \; .
\end{eqnarray}

Similarly for $G_{\Sigma^* , \Sigma}(\omega, \omega')$ we have:
\begin{equation}\label{lam6}			
4  \int_0^{\infty} dt \int dx e^{-ikx} 
\delta (-x-vt){\bf Tr} \{ \langle \pi (q)|u(0) {\bar d}(x) |0\rangle  
 [\gamma_\rho C iS^T(-x) C\gamma_{\nu}] \} \; ,
\end{equation}

To the present approximation, we need the following OPE near the light cone 
for two- and three-particle pion wave functions \cite{bely95}:
\begin{eqnarray}\label{phipi}
<\pi(q)| {\bar d} (x) \gamma_{\mu} \gamma_5 u(0) |0>&=&-i f_{\pi} q_{\mu} 
\int_0^1 du \; e^{iuqx} (\varphi_{\pi}(u) +x^2 g_1(u) + {\cal O}(x^4) ) 
\nonumber \\
&+& f_\pi \big( x_\mu - {x^2 q_\mu \over q x} \big) 
\int_0^1 du \; e^{iuqx}  g_2(u) \hskip 3 pt  , \label{ax} \\
<\pi(q)| {\bar d} (x) i \gamma_5 u(0) |0> &=& {f_{\pi} m_{\pi}^2 \over m_u+m_d} 
\int_0^1 du \; e^{iuqx} \varphi_P(u)  \hskip 3 pt ,
 \label{pscal}  \\
<\pi(q)| {\bar d} (x) \sigma_{\mu \nu} \gamma_5 u(0) |0> &=&i(q_\mu x_\nu-q_\nu 
x_\mu)  {f_{\pi} m_{\pi}^2 \over 6 (m_u+m_d)} 
\int_0^1 du \; e^{iuqx} \varphi_\sigma(u)  \hskip 3 pt .
 \label{psigma}
\end{eqnarray}
\noindent 

\begin{eqnarray}
& &<\pi(q)| {\bar d} (x) \sigma_{\alpha \beta} \gamma_5 g_s 
G_{\mu \nu}(ux)u(0) |0>=
\nonumber \\ &&i f_{3 \pi}[(q_\mu q_\alpha g_{\nu \beta}-q_\nu q_\alpha g_{\mu \beta})
-(q_\mu q_\beta g_{\nu \alpha}-q_\nu q_\beta g_{\mu \alpha})]
\int {\cal D}\alpha_i \; 
\varphi_{3 \pi} (\alpha_i) e^{iqx(\alpha_1+v \alpha_3)} \;\;\; ,
\label{p3pi} 
\end{eqnarray}

\begin{eqnarray}
& &<\pi(q)| {\bar d} (x) \gamma_{\mu} \gamma_5 g_s 
G_{\alpha \beta}(vx)u(0) |0>=
\nonumber \\
&&f_{\pi} \Big[ q_{\beta} \Big( g_{\alpha \mu}-{x_{\alpha}q_{\mu} \over q \cdot 
x} \Big) -q_{\alpha} \Big( g_{\beta \mu}-{x_{\beta}q_{\mu} \over q \cdot x} 
\Big) \Big] \int {\cal{D}} \alpha_i \varphi_{\bot}(\alpha_i) 
e^{iqx(\alpha_1 +v \alpha_3)}\nonumber \\
&&+f_{\pi} {q_{\mu} \over q \cdot x } (q_{\alpha} x_{\beta}-q_{\beta} 
x_{\alpha}) \int {\cal{D}} \alpha_i \varphi_{\|} (\alpha_i) 
e^{iqx(\alpha_1 +v \alpha_3)} \hskip 3 pt  \label{gi} 
\end{eqnarray}
\noindent and
\begin{eqnarray}
& &<\pi(q)| {\bar d} (x) \gamma_{\mu}  g_s \tilde G_{\alpha \beta}(vx)u(0) |0>=
\nonumber \\
&&i f_{\pi} 
\Big[ q_{\beta} \Big( g_{\alpha \mu}-{x_{\alpha}q_{\mu} \over q \cdot 
x} \Big) -q_{\alpha} \Big( g_{\beta \mu}-{x_{\beta}q_{\mu} \over q \cdot x} 
\Big) \Big] \int {\cal{D}} \alpha_i \tilde \varphi_{\bot}(\alpha_i) 
e^{iqx(\alpha_1 +v \alpha_3)}\nonumber \\
&&+i f_{\pi} {q_{\mu} \over q \cdot x } (q_{\alpha} x_{\beta}-q_{\beta} 
x_{\alpha}) \int {\cal{D}} \alpha_i \tilde \varphi_{\|} (\alpha_i) 
e^{iqx(\alpha_1 +v \alpha_3)} \hskip 3 pt . \label{git} 
\end{eqnarray}
\noindent 
The operator $\tilde G_{\alpha \beta}$  is the dual of $G_{\alpha \beta}$:
$\tilde G_{\alpha \beta}= {1\over 2} \epsilon_{\alpha \beta \delta \rho} 
G^{\delta \rho} $; ${\cal{D}} \alpha_i$ is defined as 
${\cal{D}} \alpha_i =d \alpha_1 
d \alpha_2 d \alpha_3 \delta(1-\alpha_1 -\alpha_2 
-\alpha_3)$. 
Due to the choice of the
gauge  $x^\mu A_\mu(x) =0$, the path-ordered gauge factor
$P \exp\big(i g_s \int_0^1 du x^\mu A_\mu(u x) \big)$ has been omitted.

The wave function $\varphi_{\pi}(u)$ is associated with the leading twist 2 
operator, $g_1(u)$ and $g_2(u)$ correspond to twist 4 operators, and $\varphi_P(u)$ and 
$\varphi_\sigma (u)$ to twist 3 ones. 
The function $\varphi_{3 \pi}$ is of twist three, while all the wave 
functions appearing in eqs.(\ref{gi}), (\ref{git}) are of twist four.
The wave functions $\varphi (x_i,\mu)$ ($\mu$ is the renormalization point) 
describe the distribution in longitudinal momenta inside the pion, the 
parameters $x_i$ ($\sum_i x_i=1$) 
representing the fractions of the longitudinal momentum carried 
by the quark, the antiquark and gluon.

The wave function normalizations immediately follow from the definitions
(\ref{phipi})-(\ref{git}):
$\int_0^1 du \; \varphi_\pi(u)=\int_0^1 du \; \varphi_\sigma(u)=1$,
$\int_0^1 du \; g_1(u)={\delta^2/12}$,
$\int {\cal D} \alpha_i \varphi_\bot(\alpha_i)=
\int {\cal D} \alpha_i \varphi_{\|}(\alpha_i)=0$,
$\int {\cal D} \alpha_i \tilde \varphi_\bot(\alpha_i)=-
\int {\cal D} \alpha_i \tilde \varphi_{\|}(\alpha_i)={\delta^2/3}$,
with the parameter $\delta$ defined by 
the matrix element: 
$<\pi(q)| {\bar d} g_s \tilde G_{\alpha \mu} \gamma^\alpha u |0>=
i \delta^2 f_\pi q_\mu$.

Expressing (\ref{lam5}) and (\ref{lam6}) with the pion wave 
functions, we arrive at:
\begin{eqnarray}\label{quark1}\nonumber
G_{\Sigma^* , \Lambda}(\omega, \omega')= 
{i\over 3}f_\pi\int_0^{\infty} dt 
\int_0^1 du e^{i (1-u) {\omega t \over 2}}
e^{i u {\omega' t \over 2}} 
\{
{1\over \pi^2 t^2} \mu_{\pi}\varphi_{\sigma}(u)&\\ 
-(\langle {\bar q} q \rangle +{t^2\over 16} 
\langle {\bar q}g_s\sigma\cdot G q \rangle )
[\varphi_\pi (u)+{it\over q\cdot v} g_2(u) +t^2 g_1(u)]
 \} &\\
+{2\over \pi^2}f_{3\pi} \int {dt \over t} \int_0^1du u 
\int {\cal D} \alpha_i 
e^{i {\omega t \over 2}[1-(\alpha_1 -u\alpha_3)]}
e^{i {\omega' t \over 2}(\alpha_1 -u\alpha_3)}
(q\cdot v) \varphi_{3\pi} (\alpha_i) &  \;,
\end{eqnarray}
\begin{eqnarray}\label{quark2}\nonumber
G_{\Sigma^* , \Sigma}(\omega, \omega')= 
{2\over \pi^2}f_\pi\int_0^{\infty} {dt \over t^3}
\int_0^1 du e^{i (1-u) {\omega t \over 2}}
e^{i u {\omega' t \over 2}} 
[\varphi_\pi (u)+{it\over q\cdot v} g_2(u) +t^2 g_1(u)]
 \} &\\
+{f_\pi\over \pi^2} \int {dt \over t} \int_0^1du  
\int {\cal D} \alpha_i 
e^{i {\omega t \over 2}[1-(\alpha_1 -u\alpha_3)]}
e^{i {\omega' t \over 2}(\alpha_1 -u\alpha_3)} 
[\tilde \varphi_{\|}(\alpha_i)-\tilde \varphi_\bot(\alpha_i)
+({1\over 2}-u)\varphi_{\|}(\alpha_i)] &  \;,
\end{eqnarray}
where $\mu_{\pi}=1.76$GeV, 
$f_{\pi}=132$MeV, $\langle {\bar q} q \rangle=-(225\mbox{MeV})^3$, 
$\langle {\bar q}g_s\sigma\cdot G q \rangle =m_0^2\langle {\bar q} q \rangle$, 
$m_0^2=0.8$GeV$^2$. 
For large euclidean values of $\omega$ and $\omega'$ 
this integral is dominated by the region of small $t$, therefore it can be 
approximated by the first a few terms.

After Wick rotations and making double Borel transformation 
with the variables $\omega$ and $\omega'$
the single-pole terms in (\ref{pole}) are eliminated. 
Subtracting the continuum contribution which is modeled by the 
dispersion integral in the region 
$\omega ,\omega' \ge \omega_c$, we arrive at:
\begin{eqnarray}\label{final-a}
\nonumber
 g_3 f_{\Sigma^*} f_{\Lambda} =& -{f^2_{\pi}\over 8\sqrt{3}\pi^2}
 e^{ {\Lambda_{\Sigma^*} +\Lambda_{\Lambda}  \over T }}
 \{
\mu_{\pi}\varphi_{\sigma}(u_0) T^3 f_2({\omega_c\over T})
-a [\varphi_\pi (u_0)T f_0({\omega_c\over T})\\ 
&-{4\over T}g_1(u_0) +{4\over T}G_2(u_0)]
+{am_0^2\over 4T} [\varphi_\pi (u_0) 
-{4\over T^2}g_1(u_0) +{4\over T^2}G_2(u_0)] \} \\
&
+{f_{\pi}f_{3\pi}\sqrt{3}\over 4\pi^2} I^G_3 (u_0) T^3f_3({\omega_c\over T})
\;,
\end{eqnarray}
where $f_n(x)=1-e^{-x}\sum\limits_{k=0}^{n}{x^k\over k!}$ is the factor used 
to subtract the continuum, $\omega_c$ is the continuum threshold.
$u_0={T_1 \over T_1 + T_2}$, 
$T\equiv {T_1T_2\over T_1+T_2}$, $T_1$, $T_2$ are the Borel parameters
$a=-(2\pi )^2 \langle {\bar q}q \rangle$.  The functions $G_2(u_0)$ 
and $I^G_3 (u_0)$ are defined as:
\begin{equation}
G_2 (u_0)=\int_0^{u_0} g_2(u)du \; ,
\end{equation}
\begin{equation}
I^G_3 (u_0) =\int_{u_0}^{ {1+u_0\over 2}} d\alpha_1 
[ {\varphi_{3\pi} (\alpha_1, 1+u_0-2\alpha_1 ,\alpha_1 -u_0 )\over 
\alpha_1 -u_0 } - 
\int_0^{1-\alpha_1} d\alpha_3 {\varphi_{3\pi} (\alpha_1, 1-\alpha_1-\alpha_3, 
\alpha_3 )\over \alpha_3^2} ] \; .
\end{equation}

We have used integration by parts 
to absorb the factors $(q\cdot v)$ and $1/(q\cdot v)$. In this way we 
arrive at the simple form after double Borel transformation. 
In obtaining (\ref{final-a}) we have used the Borel transformation formula:
${\hat {\cal B}}^T_{\omega} e^{\alpha \omega}=\delta (\alpha -{1\over T})$.

Similarly we have:
\begin{eqnarray}\label{final-b}\nonumber
 g_2 f_{\Sigma^*} f_{\Sigma} = &-{3\sqrt{2}\over 8\pi^2}f^2_{\pi}
 e^{{ 2\Lambda_{\Sigma} \over T }}
 \{ \varphi_{\pi}(u_0)T^4f_3({\omega_c\over T})
+4G_2(u_0)T^2 f_1({\omega_c\over T})
\\ 
&-4g_1(u_0)T^2 f_1({\omega_c\over T})
-{a\over 9}\mu_\pi \varphi_\sigma (u_0) 
+{am_0^2\over 36T^2}\mu_\pi \varphi_\sigma (u_0) \} \\
&
+{3\sqrt{2}\over 4\pi^2}f^2_{\pi} I_4^G (u_0) T^2f_1({\omega_c\over T})
\;,
\end{eqnarray}
where the function $I^G_4 (u_0)$ is defined as:
\begin{equation}
I^G_4 (u_0) =\int_{u_0}^{ {1+u_0\over 2}} d\alpha_1 
\int_0^{1-\alpha_1} {d\alpha_3\over \alpha_3} 
[\tilde \varphi_{\|}(\alpha_i)-\tilde \varphi_\bot(\alpha_i)
+({1\over 2}-{\alpha_1 -u_0\over \alpha_3} )\varphi_{\|}(\alpha_i)] 
\; .
\end{equation}

From (\ref{final-a}) and (\ref{final-b}) we know that both $g_2$ and $g_3$
are negative using the notations in \cite{cho}. 
In the following we will always discuss the absolute values of $g_{2,3}$.
%%%%%%%%%%%%%%%%%%%%%%%%%%%%%%%%%%%%%%%%%%%%%%%%%%%%%%%%%%%%%%%%%%%%%%%%%%%%%%%%

\section{Determination of the parameters}

\label{sec3} 
In order to obtain the coupling constants 
from (\ref{final-a})-(\ref{final-b}) we need 
the mass parameters $\bar\Lambda$'s and the coupling constants $f$'s of the
corresponding interpolating currents as input. 
The results are \cite{mass}
\begin{eqnarray}
\label{fvalue}
&&\bar\Lambda_{\Lambda}=0.8 ~~\mbox{GeV}\hspace{1.2cm}f_{\Lambda}
=(0.018\pm 0.002) ~~\mbox{GeV}^3\;,\nonumber\\
&&\bar\Lambda_{\Sigma}=1.0 ~~\mbox{GeV}\hspace{1.1cm}f_{\Sigma}
=(0.04\pm 0.004)  ~~\mbox{GeV}^3\;.
\end{eqnarray}
For the sum rule (\ref{final-a}) and (\ref{final-b}) 
the continuum threshold is $\omega_c =(2.5\pm 0.1)$GeV.

We use the wave functions adopted in \cite{bely95} to compute 
the coupling constants.
Moreover, we choose to work at the symmetric point $T_1=T_2=2T$, i.e., 
$u_0 ={1\over 2}$ as traditionally done in literature \cite{bely95}. Such a
choice is very reasonable for the sum rules for $g_2$ since $\Sigma_c^*$ and 
$\Sigma_c$ are degenerate in the leading order of HQET. 
The mass difference between $\Sigma_c^*$ and $\Lambda_c$ is about $0.2$GeV.
Due to the large values of $T_1$, $T_2\sim  
3.2 \mbox{GeV} \gg \Delta $ used below, the choice of $T_1=T_2$ is acceptable.
We use the scale $\mu =1.3$GeV, at which 
the values of the various functions appearing 
in (\ref{final-a})-(\ref{final-b}) at $u_0={1\over2}$ are: 
$\varphi_\pi(u_0)=1.22$,
$\varphi_P(u_0)=1.142$, 
$\varphi_\sigma(u_0)=1.463$, 
$g_1(u_0)=0.034 $GeV$^2$,
$G_2(u_0)=-0.02 $GeV$^2$,
$I^G_3 (u_0)=-2.75$ and
$I^G_4 (u_0)=-0.24 $GeV$^2$. We have used the asymptotic forms for 
the wave functions $\varphi_{3\pi}(\alpha_i)$, 
$\varphi_\bot(\alpha_i)$, $\varphi_{\|}(\alpha_i)$,
$\tilde \varphi_\bot(\alpha_i)$ and 
$\tilde \varphi_{\|}(\alpha_i)$ to calculate 
$I^G_3 (u_0)$ and $I^G_4 (u_0)$, since these wave functions are not 
known very well. $f_{3\pi}=0.0035$GeV$^2$.
%%%%%%%%%%%%%%%%%%%%%%%%%%%%%%%%%%%%%%%%%%%%%%%%%%%%%%%%%%%%%%%%%%%%%%%%%%%%%%%

\section{Numerical results and discussion}

\label{sec4}

We now turn to the numerical evaluation of the sum rules for the coupling
constants. 
Since the spectral density of the sum rule (\ref{final-a})-(\ref{final-b}) 
$\rho (s)$ is either proptional to $s^2$ or $s^3$, the continuum has to be 
subtracted carefully. We use the value of the continuum theshold $\omega_c$ 
determined from the corresponding mass sum rule at the leading order of 
$\alpha_s$ and $1/m_Q$ \cite{mass}.

The lower limit of $T$ is 
determined by the requirement that the terms of higher
twists in the operator expansion is reasonably smaller than the
leading twist, say $\leq 1/3$ of the latter. This leads to $T>1.3$ GeV for
the sum rules (\ref{final-a})-(\ref{final-b}).
In fact the twist-four terms contribute only a few percent to the sum rules. 
The upper limit of $T$ is constrained by the requirement that
the continuum contribution is less than $50\%$. This corresponds to $T<2.2$GeV.

The variation of $g_{2,3}$ with the Borel parameter $T$ and $\omega_c$ is 
presented in Fig. 1 and Fig. 2. The curves correspond to 
$\omega_c =2.4, 2.5, 2.6$GeV from bottom to top respectively.
Stability develops for the sum rules (\ref{final-a}) and (\ref{final-b}) 
in the region $1.3$ GeV $<$$T$$<$$2.2$ GeV, we get: 
\begin{eqnarray}
\label{result}
 &&g_2 f_{\Sigma^*} f_{\Sigma} =(2.5\pm 0.4)\times 10^{-3}\mbox{GeV}^6\;,\\
 &&g_3 f_{\Sigma^*} f_{\Lambda}=(6.8\pm 0.4)\times 10^{-4}\mbox{GeV}^6\;,
\end{eqnarray}
where the errors refers to the variations with $T$ and 
$\omega_c$ in this region. And the central value corresponds 
to $T=1.6$GeV and $\omega_c =2.5$GeV.

Combining (\ref{fvalue}) we arrive at 
\begin{eqnarray}
\label{final}
 &&g_2  =1.56\pm 0.3\pm 0.3\;,\\
 &&g_3 =0.94\pm 0.06\pm 0.2\;,
\end{eqnarray}
where the second error takes into account the uncertainty in f's.

The recent CLEO measurement \cite{cleo} 
of the $\Sigma_c^* \to \Lambda_c \pi$ decays gives 
$g_3=\sqrt{3}(0.57\pm 0.1)$ \cite{g3}, where the factor $\sqrt{3}$ arises from 
the different notaions.
The decay $\Sigma_c^* \to \Sigma_c \pi $ is kinematically forbidden so the  
direct measurement of $g_2$ is impossible. 
In the large-$N_c$ limit of QCD ($N_c$ is the number of colors) $g_{2,3}$ is related 
to the nucleon axial charge $g_A=1.25$, $g_2=3/2g_A$, $g_3=\sqrt{3/2}g_A$ 
\cite{gura93,jenkins93}. 
The quark model result is $g_3=\sqrt{3}\times 0.61$, 
$g_2=1.5\times (0.93\pm 0.16)$ \cite{g3}. 
The short-distance QCD sum rules with the external field method \cite{grozin97}
yields $g_2=1.5\times (0.4\sim 0.7)$, $g_3=\sqrt{3/2}\times (0.4\sim 0.7)$.
In this work we employ light-cone QCD sum rules to calculate the strong 
coupling constants $g_{2,3}$. Both sum rules for $g_2$ and $g_3$ is very 
stable with reasonable variations of the Borel parameter $T$
and the continuum threshold $\omega_c$ as can be seen from Fig. 1 and Fig. 2. 
It is interesting to note that the numerical values of $g_2$ and $g_3$ 
are consistent with both the experimental data and the quark model result \cite{g3}.
It is also interesting to notice the deviation from the large $N_c$ limit prediction
\cite{gura93,jenkins93}. Moreover the result from the short-distance QCD sum rules 
is compatible with the present work with the light-cone QCD sum rule approach if 
we use the same values of $\omega_c$ and $f$'s though the errors are quite large.

%%%%%%%%%%%%%%%%%%%%%%%%%%%%%%%%%%%%%%%%%%%%%%%%%%%%%%%%%%%%%%%%%%%%%%%%%%%%%%%%

\vspace{0.8cm} {\it Acknowledgements:\/} S.-L. Zhu was supported by
the National Postdoctoral Science Foundation of China and Y.D. was supported by 
the National Natural Science Foundation of China.
\bigskip
\vspace{1.cm}

{\bf Figure Captions}
\vspace{2ex}
\begin{center}
\begin{minipage}{120mm}
{\sf
Fig. 1.} \small{Dependence of $g_2$ on the Borel parameter $T$ for 
different values of the continuum threshold $\omega_c$. 
From top to bottom the curves correspond 
to $\omega_c=2.6, 2.5, 2.4$ GeV. }
\end{minipage}
\end{center}
\begin{center}
\begin{minipage}{120mm}
{\sf
Fig. 2.} \small{Dependence of $g_3$ on the Borel parameter $T$ for 
different values of the continuum threshold $\omega_c$. 
From top to bottom the curves correspond 
to $\omega_c=2.6, 2.5, 2.4$ GeV. }
\end{minipage}
\end{center}

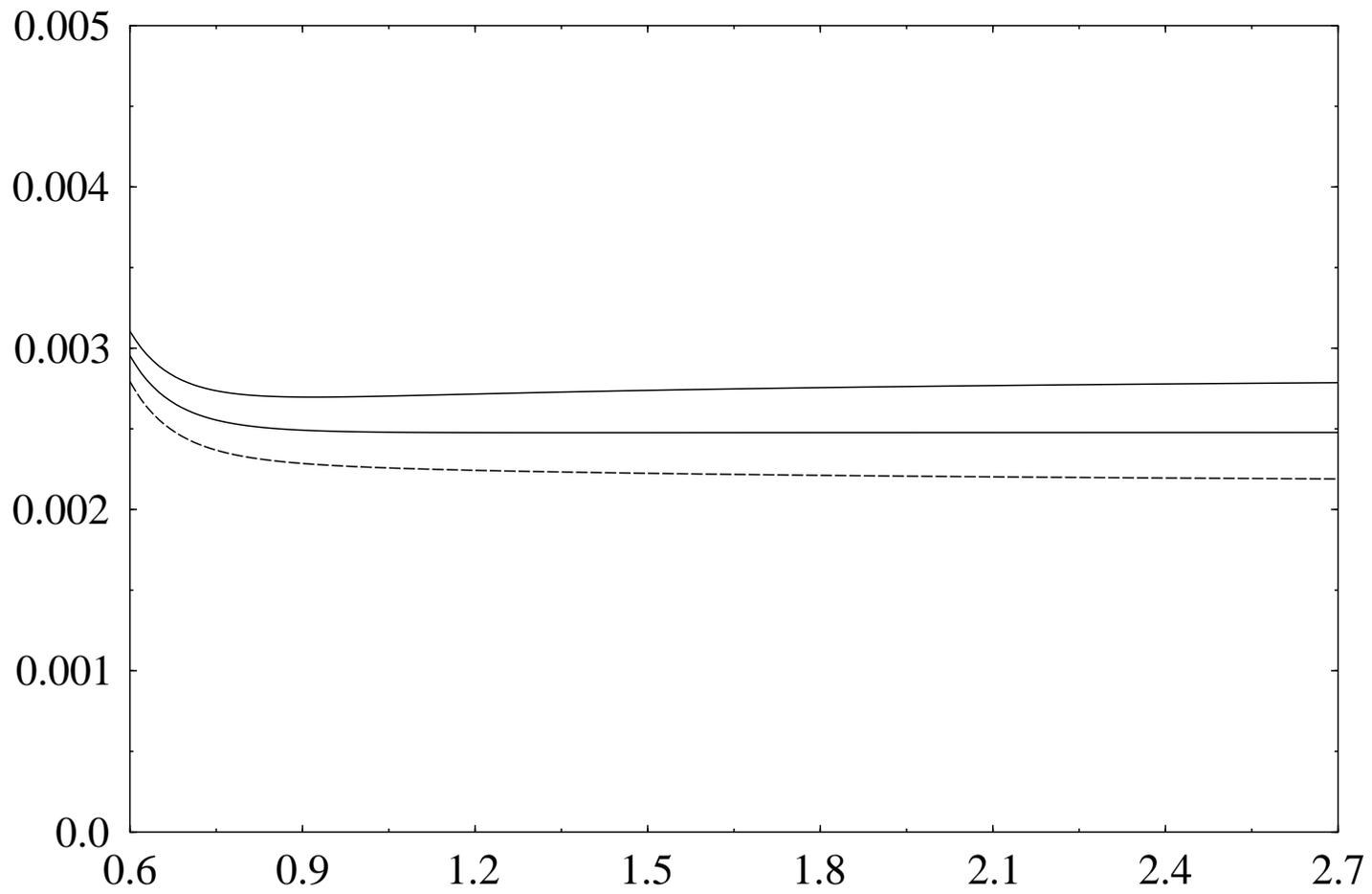
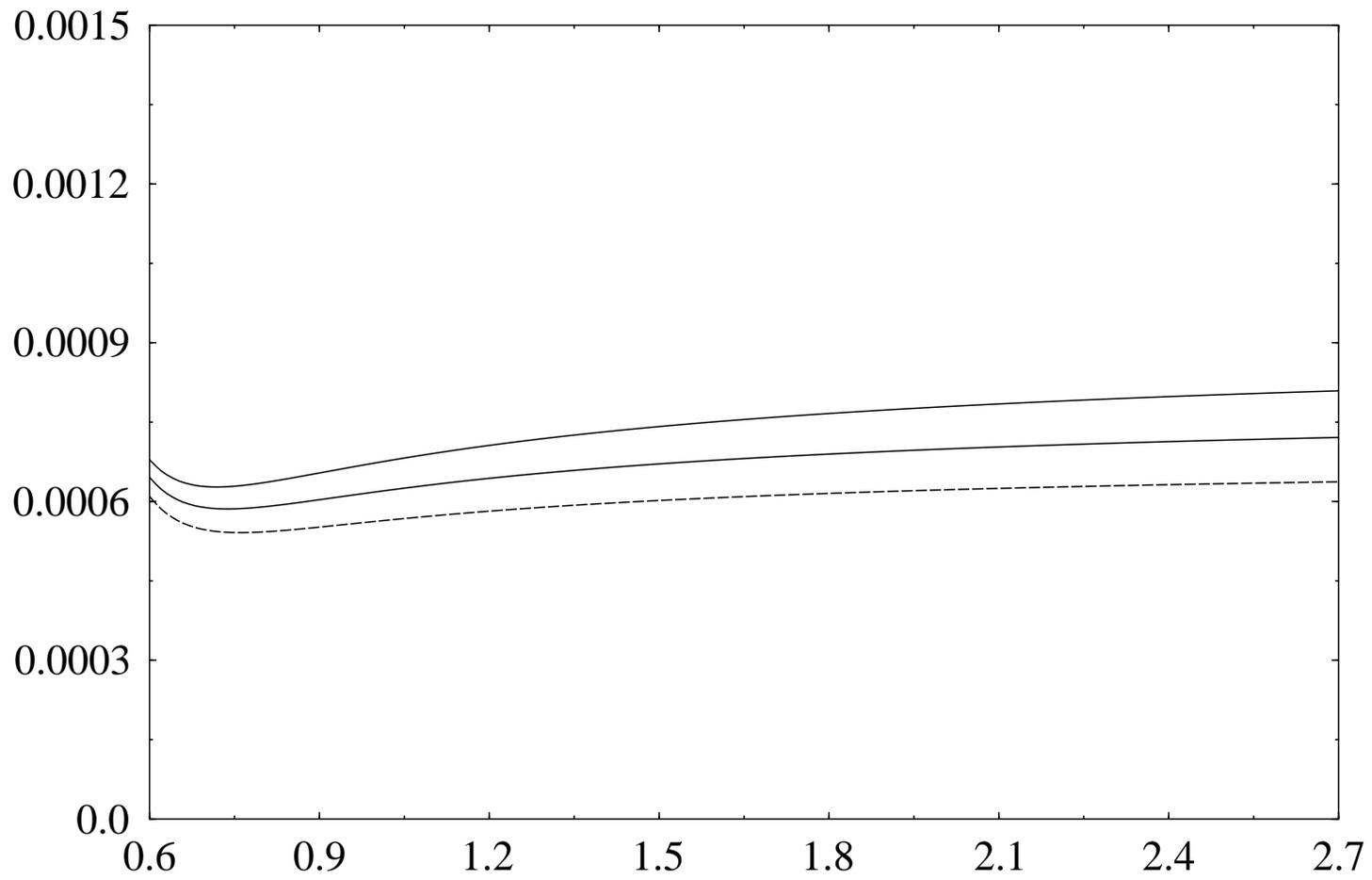
\end{document}